\documentclass[a4paper,12pt]{article}

\def\a{\alpha}
\def\b{\beta}
\def\d{\delta}
\def\la{\lambda}
\def\m{\mu}
\def\n{\nu}
\def\t{\theta}

\def\p{\partial}
\def\nn{\nonumber\\}
\newcommand{\lto}{\longrightarrow}
\newcommand{\cF}{{\cal F}}
\newcommand{\DSP}{\displaystyle}

\def\aD#1{$\overline{\mathrm{D}#1}$}

\def\Tr{\mathop{\mathrm{Tr}}\nolimits}
\def\STr{\mathop{\mathrm{STr}}\nolimits}
\def\Str{\mathop{\mathrm{Str}}\nolimits}
\newcommand\re{\mathop{\mathrm{Re}}\nolimits}
\newcommand\im{\mathop{\mathrm{Im}}\nolimits}

\def\mtx#1#2{\left(\begin{array}{@{\,}#1@{\,}}#2\end{array}\right)}
\makeatletter\@addtoreset{equation}{section}
\renewcommand{\thefigure}{\@arabic\c@figure}
\makeatother

\title{Tachyon condensation in unbalanced D$\overline{\mathrm{D}}$
	system}

\author{\textsc{Akira Ishida$^1$}\footnote{e-mail:
		ishida@skku.edu},~
	\textsc{Shozo Uehara$^2$}\footnote{e-mail:
		uehara@eken.phys.nagoya-u.ac.jp}~~and
	\textsc{Tomoki Yada$^2$}\footnote{e-mail:
		yada@eken.phys.nagoya-u.ac.jp}\\[5mm]
	\textsl{$^1$BK21 Physics Research Division and
		Institute of Basic Science,}\\
	\textsl{Sungkyunkwan University, Suwon 440-746, Korea}\\
	\textsl{$^2$Department of Physics, Nagoya University,}\\
	\textsl{Chikusa-ku, Nagoya 464-8602, Japan}\hfill}
\date{}
\begin{document}
\maketitle
\begin{flushright}
\vspace*{-85mm}
	DPNU-05-14\\
	hep-th/0601050
\end{flushright}

\vspace{55mm}
\begin{abstract}
The tachyon condensation is studied in asymmetric D\aD{} systems.
Taking a system of two pairs of D5-\aD{5} in type IIB superstring
theory in the background of large $N$ D5-branes,
we show that one BPS D1-brane comes out after the condensation.
It is also seen that the BPS D1-brane feels no force from the
background D5-branes.
We also show that the inclusion of the fluctuation fields gives an
expected Dirac-Born-Infeld (DBI) action of the resultant D1-brane.
On the other hand, in the case of one pair of D5-\aD{5} in the same
background, we show that the resultant BPS D3-brane experiences
attractive force from the background D5-branes.
\end{abstract}

\newpage

\section{Introduction}
Open string tachyon dynamics of an unstable D-brane system (e.g. a
non-BPS D-brane or brane-antibrane pair) leads us to a deeper insight
into nonperturbative aspects of string theory.
In recent years, many works have been done on the open string tachyon
condensation (for a review see Ref.\cite{Sen:04nf}).
For example, through tachyon condensation unstable D-brane systems can
produce lower-dimensional D-branes, which is called brane descent
relations.
This construction was systematized in Ref.\cite{W:98cd}, in which the
D-branes in type IIB theory are understood as the bound states of a
number of D9-\aD{9} pairs with tachyon condensation and it was shown
that D-brane charges are classified by K-theory.
Some exact results can be obtained by using the boundary string field
theory (BSFT) \cite{W:92qy}.
Actually, BSFT gives the exact form of the tachyon potential and it
was also shown that the tachyon condensation on a D9-\aD{9} system in
type IIB superstring theory gives a non-BPS D8-brane and BPS D7-brane
with their correct tensions \cite{BSFT:DDb}.
Another interesting process of the tachyon condensation is a time
dependent one.
The time dependent boundary state describing the decay of an unstable
D-brane was constructed in Refs.\cite{Sen:02nu,Sen:02in} and it was
shown that the nonzero energy density, which is called the tachyon
matter, remains after the decay.
It was also shown that open string excitations of a tachyon and a
gauge field on the tachyon matter disappear \cite{Sen:02an,IU0}.
These evidences strongly support Sen's conjecture \cite{Sens:}.

It is, of course, important to consider the tachyon condensation on
various D-branes.
In BSFT, however, it is difficult to investigate such systems except
those of some specific numbers of D-branes since, in general,  we
cannot use the boundary fermion technique straightforwardly.
Actually, in Ref.\cite{J:03} they attempted to analyze the system of
two D9-branes and one \aD{9}-brane by approximating the BSFT action.
However, for example, the correct descent relation of the D-brane
tension was not obtained and it seems difficult to improve the
approximation.

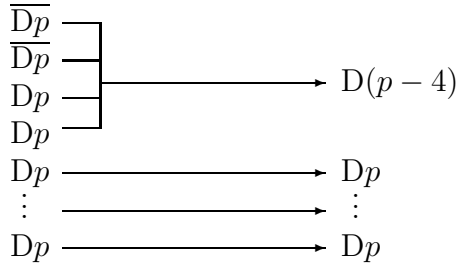
\begin{figure}[ht]
\begin{center}
\unitlength1mm
\begin{picture}(55,40)
\put(0,35){\aD{p}}
\put(0,30){\aD{p}}
\put(0,25){D$p$}
\put(0,20){D$p$}
\put(0,15){D$p$}
\put(0,10){~$\vdots$}
\put(0,5){D$p$}
\put(7,16){\vector(1,0){35}}
\put(7,11){\vector(1,0){35}}
\put(7,6){\vector(1,0){35}}
\put(44,15){D$p$}
\put(44,10){~$\vdots$}
\put(44,5){D$p$}
\put(7,36){\line(1,0){5}}
\put(7,26){\line(1,0){5}}
\put(12,28){\vector(1,0){30}}
\put(7,31){\line(1,0){5}}
\put(7,22){\line(1,0){5}}
\put(12,22){\line(0,1){14}}
\put(44,27){D($p-4$)}
\end{picture}
\caption{Schematic evolution 1.}\label{F:1}
\end{center}
\end{figure}

Let us consider $(N+2)$ D$p$-branes and two \aD{p}-branes.
This is an unstable system and the string between any pair of
D$p$-\aD{p} has a tachyonic mode.
One may expect that the tachyon condensation would occur on two pairs
of D$p$-\aD{p} to produce a D$(p-4)$-brane, and hence it would become
a BPS system of one D$(p-4)$-brane and $N$ D$p$-branes.
This condensation process is schematically depicted in Figure
\ref{F:1}.
Since there is no force between the D$(p-4)$-brane and the $N$
D$p$-branes, they can be separated. This implies that the original
tachyonic modes of the strings, especially stretched between the
\aD{p}-branes and the ``spectators'' of $N$ D$p$-branes, have
disappeared\footnote{These modes are presumably raised to be
massless modes on the D$(p-4)$-brane worldvolume.}
after the condensation of the tachyonic modes of the
strings between two pairs of D$p$-\aD{p} system.
This could be realized, however, it has not been explicitly shown so
far.

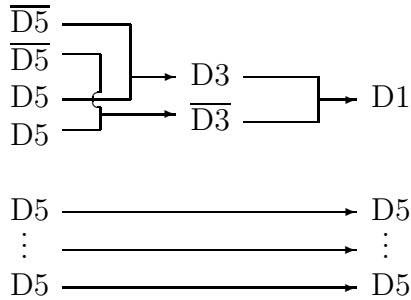
\begin{figure}[ht]
\begin{center}
\unitlength1mm
\begin{picture}(55,45)
\put(0,40){\aD{5}}
\put(0,35){\aD{5}}
\put(0,30){D5}
\put(0,25){D5}
\put(0,15){D5}
\put(0,10){~$\vdots$}
\put(0,5){D5}
\put(7,16){\vector(1,0){39}}
\put(7,11){\vector(1,0){39}}
\put(7,6){\vector(1,0){39}}
\put(48,15){D5}
\put(48,10){~$\vdots$}
\put(48,5){D5}
\put(7,41){\line(1,0){9}}
\put(7,31){\line(1,0){9}}
\put(16,31){\line(0,1){10}}
\put(16,34){\vector(1,0){6}}
\put(24,33){D3}
\put(7,37){\line(1,0){5}}
\put(7,27){\line(1,0){5}}
\put(12,31){\oval(2,2)[l]}
\put(12,32){\line(0,1){5}}
\put(12,27){\line(0,1){3}}
\put(12,29){\vector(1,0){10}}
\put(24,27){\aD{3}}
\put(31,34){\line(1,0){10}}
\put(31,28){\line(1,0){10}}
\put(41,28){\line(0,1){6}}
\put(41,31){\vector(1,0){5}}
\put(48,30){D1}
\end{picture}
\caption{Schematic evolution 2.}\label{F:2}
\end{center}
\end{figure}

In this paper we shall attempt to analyze the tachyon condensation on
such a system of multi D-branes, or two \aD{5}-branes and
$(N+2)$ D5-branes where two of the D5-branes together with the two
\aD{5}-branes are separated from the rest of the $N$ D5-branes.
We consider the large $N$ case, which allows us to treat the $N$
D5-branes as a background geometry of the spacetime.
The two \aD{5}-branes and two of the ($N+2$) D5-branes condensate each
other, which produces one D3-\aD{3} pair. Further condensation may
occur on the D3-\aD{3} and eventually one D1-brane is left stably in
the D5-brane background.
Figure \ref{F:2} presents a cascade of the tachyon condensation
schematically.
We give the tachyon profiles explicitly which lead to various lower
dimensional D-branes after the condensation.
Especially, we shall show that one BPS D1-branes is obtained, in which
there is no force between the resultant one D1-brane and the $N$
D5-branes and the total energy is independent of the distance between
them.
On the other hand, there is attractive force between a D3-brane and a
D5-brane. Then, we also consider the tachyon condensation on a pair
of D5-\aD{5} in the D5-brane background. We show that unlike the
codimension 4 case the resultant BPS D3-brane experiences attractive
force from the D5-brane background.

The organization of the paper is as follows.
In section \ref{S:2}, we first analyze the system by using a
non-Abelian Dirac-Born-Infeld (DBI) action of the D$p$-\aD{p} system
\cite{Ga:04, Sen:03}.
We consider the tachyon effective field theory action of pairs of
D$p$-\aD{p} in the D$p$-brane background.
In section \ref{S:3}, we will give the tachyon profiles explicitly
which lead to various lower dimensional D-branes.
In particular, it is possible to leave one BPS D1-brane stably
in the background of large $N$ D5-branes.
In section \ref{S:4}, we discuss the equations of motion for the
tachyon profiles given in section \ref{S:3}.
In section \ref{S:5}, taking into account the fluctuations around the
vortex, we shall have the DBI actions including transverse fields.
In section \ref{S:6}, we analyze the system in BSFT.
Finally, we conclude the paper with some discussion.

\section{Tachyon effective field theory action}\label{S:2}
In this section, we consider the tachyon effective action on
coincident $M$ pairs of D$p$-\aD{p} in the background of $N$
D$p$-branes.
Hereafter we consider in type IIB superstring theory and hence a
D$p$-brane with odd $p$ satisfies the BPS condition.
As we have explained in the introduction, it is not easy to analyze
a system of $(N+M)$ D$p$-branes and $M$ \aD{p}-branes for a general
set of ($N+M,M$).
Thus we take $N$ to be large and the distance between $M$ pairs of
D$p$-\aD{p} and the rest of $N$ D$p$-branes to be sufficiently
larger than the string scale so that the $N$ D$p$-branes produce the
background geometry of the spacetime.
The spacetime metric, the dilaton $\phi$ and the RR field
$C_{0\ldots p}$ in this background are given by
\begin{eqnarray}
 &&G_{\m\n}=H_p^{-\frac{1}{2}}\,\eta_{\m\n}\,,\quad
	G_{IJ}=H_p^{\,\frac{1}{2}}\,\d_{IJ}\,,\quad
	( 0\le \m,\n \le p,\quad p+1\le I,J \le 9 )\nn
 &&e^{-\phi}=H_p^{\frac{p-3}{4}}\,,\quad C_{0\ldots p}=H_p^{-1}\,,
  \quad H_p=1+N g_s\,\left(\frac{\sqrt{\a'}}{R_\perp}\right)^{7-p}
	\,,\label{eq:Gmn}
\end{eqnarray}
where $H_p$ is the harmonic function describing the $N$ D$p$-branes
and $R_\perp$ is the distance from the stack of the $N$ D$p$-branes.
Here we consider the region, $\sqrt{\a'}\ll R_\perp$, where the
description of this background is good and thus we can appropriately
use the DBI action in terms of this background.\footnote{Some works
have been done on the brane dynamics in the D-brane and NS5-brane
background. Kutasov studied the D-brane dynamics near NS5-branes
\cite{Ku:04a}. Following this, the D-brane dynamics in the D-brane
background was considered in \cite{KLP}.}

Now that we know the background geometry, we shall consider the
effective action.
Myers \cite{Myrs} proposed the world-volume action for $N$ coincident
D$p$-branes where the world-volume theory involves a $U(N)$ gauge
theory. The extension to a D$p$-\aD{p} system is given in
Refs.\cite{Ga:04,Sen:03}.

We consider the situation where $M$ pairs of D$p$-\aD{p} are in the
background geometry of eq.(\ref{eq:Gmn}).
This system is unstable and tachyonic modes appear between D$p$-\aD{p}
due to the opposite GSO projection.
Following Refs.\cite{Ga:04, Sen:03}, we write down our action for
the $M$ pairs of D$p$-\aD{p} as
\begin{equation}
	S_{\rm{DBI}} = -2\int\! d^{p+1}\xi\ \STr
	\left(e^{-\phi}\,V(T) \sqrt{-\det(P[G]_{\m\n}
	+ \la \textrm{S}[(\p_\m T)^\dagger (\p_\n T)])}\,\right),
\end{equation}
where $\la = 2\pi\a'$ and the symbols STr and $P$ stand for the
symmetrized trace and the pull-back of the spacetime metric,
respectively,
\begin{equation}
  P[G]_{\m\n} \equiv G_{\m\n} + 4\pi\a'\,G_{I(\m}\p_{\n)}\Phi^I
	+ (2\pi\a')^2\,G_{IJ}\,\p_\m \Phi^I \p_\n \Phi^J,
\end{equation}
where $\Phi^I$ are the spacetime coordinates of the D$p$-branes in
the static gauge.
And we have set both the Kalb-Ramond field and the field strength to
be zero, $B_{MN}=F_{\m\n}=0$, for simplicity and defined
\begin{equation}
  \textrm{S}[(\p_\m T)^\dagger (\p_\n T)] \equiv
  \frac{1}{2}(\p_\m T)^\dagger (\p_\n T)
	+ \frac{1}{2}(\p_\n T)^\dagger (\p_\m T)\,.
\end{equation}
Hereafter we ignore the transverse fluctuation of the branes
except in section \ref{S:5},
$\Phi^I=0$, and hence $P[G]_{\m\n}=G_{\m\n}$.
Note that the tachyon field\footnote{Throughout this paper tachyon
field $T$ is dimensionless.}
is described by a complex $M\times M$
matrix and in this case the tachyon potential term $V(T)$, whose
functional form is not specified explicitly here, is assumed to be a
function of $T^\dagger T$. Thus we symbolically write it as $V(|T|^2)$
henceforth. And it should damp at infinity,
$V(\infty)=0$, and is normalized by $V(0)=T_p$ where $T_p$ denotes the
tension of a BPS D$p$-brane,
\begin{equation}
	T_p = \frac{2\pi}{g_s (2\pi\sqrt{\a'})^{p+1}}\,.
\end{equation}

\section{\boldmath Tachyon condensation on 2D5-2\aD{5} with
$N$D5}\label{S:3}
In this section, we analyze specifically the tachyon condensation on
two pairs of D5-\aD{5} in the background of large $N$ D5-branes.
We give the explicit forms of the tachyon field obeying the equation of
motion and show that the expected types of D-branes are produced after
the condensation.
Actually, we shall leave the discussion on the equations of motion
till the next section and give the various tachyon profiles which
lead to a BPS D1-brane, a non-BPS D2-brane, a pair of BPS
D3-\aD{3} branes and two non-BPS D4-branes, respectively.

\subsection{Codimension 4: BPS D1-brane}
First we consider a four-dimensional vortex solution of the tachyon
field on two pairs of D5-\aD{5} where the tachyon field is expressed
by a complex $2\times 2$ matrix.
We shall see that one BPS D1-brane is produced in the large $N$
D5-brane background after the tachyon condensation.
In what follows, we expand the complex $2\times 2$ matrix tachyon
field in terms of the quaternion bases,
\begin{equation}
  Q_0= \mtx{cc}{1 & 0 \\ 0 & 1 },\quad
  Q_1= \mtx{cc}{i & 0 \\ 0 & -i},\quad
  Q_2= \mtx{cc}{0 & 1 \\ -1 & 0} ,\quad
  Q_3= \mtx{cc}{0 & i\\ i & 0}.
\end{equation}
We take a four-dimensional polar coordinate $(r,\t_1,\t_2,\t_3)$
for the four spatial directions, i.e., 2nd-, 3rd-, 4th- and
5th-directions, on the D5-\aD{5} pairs,
\begin{equation}
 \eta_{\m\nu}\,d\xi^\m d\xi^\n=
  \mtx{c}{d\xi^0\\ d\xi^1\\ dr\\ d\t_1\\ d\t_2\\ d\t_3}^T
  \mtx{ccccc@{}c}{-1 & & & & &\hbox{\Large0}\\
		& 1 & & & & \\
		& & 1 & & &\\
		& & & r^2& &\\
		& & & & r^2\sin^2\t_1 &\\
		\hbox{\Large0}& & & & &r^2\sin^2\t_1\sin^2\t_2}
  \mtx{c}{d\xi^0\\ d\xi^1\\ dr\\ d\t_1\\ d\t_2\\ d\t_3}.
\end{equation}
We take each coefficient $T_\a\ (\a=q_0,q_1,q_2,q_3)$ as follows,
\begin{eqnarray}
	T_{q_0} &=& f(ur) \cos\t_1\,,\\
	T_{q_1} &=& f(ur) \sin\t_1 \cos\t_2\,,\\
	T_{q_2} &=& f(ur) \sin\t_1 \sin\t_2 \cos\t_3\,,\\
	T_{q_3} &=& f(ur) \sin\t_1 \sin\t_2 \sin\t_3\,,
\end{eqnarray}
where $f(x)$ satisfies
\begin{equation}
    f(0) = 0,\quad f'(x)>0\quad (x \ge 0)\,,
	\quad f(\infty)=\infty,\quad f'(\infty)>0\,, \label{eq:cndf}
\end{equation}
and $u$ is a constant taken to be $\infty$ at the end,
which gives a singular vortex solution.
Then the four-dimensional vortex profile of the tachyon field is
given by
\begin{eqnarray}
  T(r,\t_1,\t_2,\t_3) &=& T_{q_0}\,Q_0 + T_{q_1}\,Q_1
	+ T_{q_2}\,Q_2 + T_{q_3}\,Q_3\nn
  &=& \mtx{cc}{
    f\cos\t_1 +if\sin\t_1 \cos\t_2 & f\sin\t_1 \sin\t_2\,e^{i\t_3}\\
    -f\sin\t_1 \sin\t_2\,e^{-i\t_3} & f\cos\t_1 -if\sin\t_1 \cos\t_2}.
  \label{eq:TPd1}
\end{eqnarray}

Now we compute the DBI action by using the above profile
(\ref{eq:TPd1}). The kinetic term of the tachyon field becomes
\begin{eqnarray}
 &&\hspace{-12mm}\textrm{S}[(\p_\m T)^\dagger (\p_\n T)]=\nn
 &&\hspace{-10mm} \mtx{cccccc}{
	0 & & & & &\\
	& 0 & & & &\hbox{\LARGE 0} \\
	& & u^2 f'^2(ur)\,Q_0 & & &\\
	& & & f^2(ur)\,Q_0 & & \\
	& & & & f^2(ur) \sin^2\t_1\,Q_0 &\\
	&\hbox{\LARGE0} & & & & f^2(ur) \sin^2\t_1 \sin^2\t_2\,Q_0},
\end{eqnarray}
and the induced metric on the world-volume is given by
\begin{equation}
  P[G]_{\m\n}=H_p^{-\frac{1}{2}} \mtx{ccccc}{
	\eta_{\a\b}Q_0 & & & &\\
	& Q_0 & & &\hbox{\Large0}\\
	& & r^2\,Q_0 & &\\
	& & & r^2 \sin^2\t_1\,Q_0 &\\
	&\hbox{\Large0} & & & r^2 \sin^2 \t_1 \sin^2\t_2\,Q_0},
\end{equation}
where $\eta_{\a\b}=\mathrm{diag}(-1,1)$.
Thus we have
\begin{eqnarray}
 &&\textrm{S}[P[G]_{\m\n} +2\pi\a' (\p_\m T)^\dagger(\p_\n T)]\nn
 &&=\mathrm{diag} \Bigl(H_p^{-\frac{1}{2}} \eta_{\a\b}Q_0,\,
    (H_p^{-\frac{1}{2}}+2\pi\a' u^2 f'^2)Q_0,\,
    (H_p^{-\frac{1}{2}} r^2 + 2\pi\a' f^2) Q_0,\,\nn
 &&\hspace{8ex} (H_p^{-\frac{1}{2}} r^2+ 2\pi\a' f^2)\sin^2\t_1 Q_0,\,
    (H_p^{-\frac{1}{2}}r^2 + 2\pi\a' f^2)\sin^2\t_1\sin^2\t_2Q_0
	\Bigr)\,.\quad\label{eq:co4det}
\end{eqnarray}
By using eq.(\ref{eq:co4det}) and collecting the most singular part of
the action in the $u\to\infty$ limit, we have
\begin{eqnarray}
 &&\hspace{-7ex}2\int\! d^6\xi\ \STr \left(e^{-\phi} V(|T|^2)
	\sqrt{-\det(P[G]_{\m\n}
	+ 2\pi\a' (\p_\m T)^\dagger (\p_\n T))}\,\right)\nn
 &\lto& 4\int\! d^6\xi\ V(|T|^2) H_p^{\frac{1}{2}}
	\sqrt{H_p^{-1} (2\pi\a')^4 u^2 f'^2(ur) f^6(ur)
	\sin^4\t_1 \sin^2\t_2}\nn
 &=& 4\cdot (2\pi\a')^2 H_p^{\frac{1}{2}} H_p^{-\frac{1}{2}}
	\int_0^\infty \! dr\ u f'(ur) f^3(ur) V(f^2(ur))
	\int\! d\Omega_4 \int\! d^2\xi\nn
 &=& 4\cdot (2\pi\a')^2 \cdot (2\pi^2) \left(\int_0^\infty
	\!dy\,y^3V(y^2)\right)\int\! d^2\xi,\quad (y\equiv f(ur)\,)\nn
 &=&(2\pi\sqrt{\a'})^4\left(2\int_0^\infty\!dy\,y^3\,V(y^2)\right)
	\int\! d^2\xi\,.\label{eq:bpsd1}
\end{eqnarray}
Notice that the $H_p$ dependence disappears in eq.(\ref{eq:bpsd1})
\cite{KLP}, which indicates that the two-dimensional object remains
stable.
Once we postulate that the constant value of the integral is given by
\begin{equation}
	2\int_0^\infty \!dy\ y^3\, V(y^2) = T_p \,,\label{eq:co4}
\end{equation}
we have, in fact, a BPS D1-brane with the correct value of the tension
$T_{p-4}= (2\pi\sqrt{\a'})^4\,T_p$\,.
In the next subsection we shall give other tachyon profiles which lead
to the lower codimension D-branes, all of which, however, are unstable
due to the attractive force from the background.

\subsection{Lower codimension branes}
First, considering a three-dimensional vortex solution, we shall get
one non-BPS D2-brane.
We take $T_{q_3}=0$ and others are given, in terms of the
three-dimensional polar coordinates, by
\begin{eqnarray}
	T_{q_0} &=& f(ur) \cos\t_1\,,\\
	T_{q_1} &=& f(ur) \sin\t_1 \cos\t_2\,,\\
	T_{q_2} &=& f(ur) \sin\t_1 \sin\t_2\,,
\end{eqnarray}
and hence the tachyon profile is
\begin{eqnarray}
  T(r,\t_1,\t_2) &=& T_{q_0}\,Q_0 + T_{q_1}\,Q_1 + T_{q_2}\,Q_2\nn
  &=& \mtx{cc}{f\cos\t_1 +if\sin\t_1 \cos\t_2 & f\sin\t_1 \sin\t_2\\
    -f\sin\t_1 \sin\t_2 & f\cos\t_1 -if\sin\t_1 \cos\t_2}\,.
	\label{eq:TPd2}
\end{eqnarray}
Thus, the most singular part of the action in the $u\to\infty$ limit
is
\begin{eqnarray}
 &&\hspace{-7ex}2\int\! d^6\xi\,\STr \left(e^{-\phi} V(|T|^2)
	\sqrt{-\det(P[G]_{\m\n}
	+ 2\pi\a' (\p_\m T)^\dagger (\p_\n T))}\right)\nn
 &\lto& 4\int\! d^6\xi\ V(|T|^2)\,H_p^{\frac{1}{2}}\,
	\sqrt{H_p^{-\frac{3}{2}}
	(2\pi\a')^3\,u^2 f'^2(ur) f^4(ur) \sin^2\t_1}\nn
 &=& \sqrt{2} (2\pi\sqrt{\a'})^3 \left(\frac{4}{\sqrt{\pi}}
	\int_0^\infty\!dy\ y^2 V(y^2)\right)
	H_p^{-\frac{1}{4}}\int\! d^3\xi\,.\label{eq:d2}
\end{eqnarray}
If we postulate
\begin{equation}
  T_p =\frac{4}{\sqrt{\pi}}
	\int_0^\infty\!dy\ y^2\,V(y^2)\,,\label{eq:co3}
\end{equation}
we have one non-BPS D2-brane.

Next, we consider another profile. $T_{q_2}=T_{q_3}=0$ and using
the two-dimensional polar coordinates, we take
\begin{eqnarray}
	T_{q_0} &=& f(ur) \cos\t_1\,,\\
	T_{q_1} &=& f(ur) \sin\t_1\,,
\end{eqnarray}
so that
\begin{equation}
  T(r,\t_1) = T_{q_0}\,Q_0 + T_{q_1}\,Q_1
  =\mtx{cc}{f\,e^{i\t_1} & 0\\ 0 & f\,e^{-i\t_1}}.\label{eq:TPd3}
\end{equation}
The most singular part of the action is
\begin{eqnarray}
 &&\hspace{-7ex}2\int\! d^6\xi\,\STr\left(e^{-\phi} V(|T|^2)
    \sqrt{-\det(P[G]_{\m\n}
	+ 2\pi\a' (\p_\m T)^\dagger (\p_\n T))}\right)\nn
 &\lto& 4\int\! d^6\xi\ V(|T|^2) H_p^{\frac{1}{2}}\,
    \sqrt{H_p^{-2}
	(2\pi\a')^2 u^2 f'^2(ur) f^2(ur)}\nn
 &=& 2 (2\pi\sqrt{\a'})^2 \left(2\int_0^\infty \!dy\ y\,V(y^2)\right)
	H_p^{-\frac{1}{2}} \int\! d^4\xi\,.\label{eq:d3}
\end{eqnarray}
This represents a pair of BPS D3-brane and BPS \aD{3}-brane once we
postulate
\begin{equation}
  2\int_0^\infty \!dy\ y\,V(y^2) = T_p\,.\label{eq:co2}
\end{equation}

Finally, we consider the codimension one case.
We set $T_{q_1}=T_{q_2}=T_{q_3}=0$ and
\begin{equation}
	T_{q_0} = f(ur)\,,
\end{equation}
which leads to the tachyon profile,\footnote{Here the coordinate $r$
takes $-\infty<r<\infty$ and hence we regard that the function $f$
satisfies $f(-x)=-f(x)$ and $f(\pm\infty)=\pm\infty$.}
\begin{equation}
  T(r) = T_{q_0}\,Q_0 = \mtx{cc}{ f & 0\\ 0 & f}.\label{eq:TPd4}
\end{equation}
Similarly, the most singular part of the action becomes
\begin{eqnarray}
 &&\hspace{-7ex}2\int\! d^6\xi\,\STr\left(e^{-\phi} V(|T|^2)
    \sqrt{-\det\left(P[G]_{\m\n} + 2\pi\a' (\p_\m T)^\dagger
	(\p_\n T)\right)}\,\right)\nn
 &\lto& 4\int\! d^6\xi\ V(|T|^2) H_p^{\frac{1}{2}}\,
    \sqrt{H_p^{-\frac{5}{2}} (2\pi\a') u^2 f'^2(ur)}\nn
 &=& 2\sqrt{2} (2\pi\sqrt{\a'})
    \left(\frac{2}{\sqrt{\pi}}\int_0^\infty\!dy\ V(y^2)\right)
	H_p^{-\frac{3}{4}} \int\! d^5\xi\,. \label{eq:d4}
\end{eqnarray}
If we postulate
\begin{equation}
  T_p = \frac{2}{\sqrt{\pi}}\int_0^\infty\!dy\ V(y^2)\,,\label{eq:co1}
\end{equation}
we have correctly two non-BPS D4-branes.
Two comments are in order: (i) In the codimension 4 case, the final
form in eq.(\ref{eq:bpsd1}) does not depend on $H_p$ at all, while
others, eqs.(\ref{eq:d2}), (\ref{eq:d3}) and (\ref{eq:d4}), depend on
$H_p$ as $H_p^{-\a}\, (\a>0)$, which indicates instability of such
D-branes, or the attractive force towards $R_\perp=0$, in the D5-brane
background.
(ii) We have not explicitly calculated the tension of the
resultant D-branes since we did not give any definite form of the
tachyon potential $V(|T|^2)$ so far.
Actually all the conditions, eqs.(\ref{eq:co4}), (\ref{eq:co3}),
(\ref{eq:co2}) and (\ref{eq:co1}), are simultaneously satisfied if we
adopt the tachyon potential of $V(|T|^2)=T_p\,e^{-T^\dagger T}$.

\section{Verification of the equation of motion}\label{S:4}
In the previous section, we have seen that the tachyon profiles
led to the expected types of D-branes. In this section we
investigate whether the tachyon profiles really obey the equations of
motion.
The Lagrangian is given by
\begin{equation}
  {\cal L} \equiv -2H_p^{\frac{1}{2}}\,\STr\left(V(|T|^2)\,
	\sqrt{-\det A}\,\right),
\end{equation}
where
\begin{equation}
  A_{\m\n}\equiv P[G]_{\m\n}
	+\la\textrm{S}[(\p_\m T)^\dagger(\p_\n T)]\,.\label{eq:A}
\end{equation}
We shall show that the equation of motion is actually satisfied.
The Euler-Lagrange equation,
\begin{equation}
 \p_\m \left(\frac{\p{\cal L}}{\p(\p_\m T^\dagger)}\right)
	- \frac{\p {\cal L}}{\p T^\dagger} = 0\,,
\end{equation}
is given by
\begin{equation}
  \la\,\p_\m\left\{\p_\n T\,V(|T|^2)\,(A^{-1})^{\m\n}
	\sqrt{-\det A}\,\right\}- 2T\,V'(|T|^2)\sqrt{-\det A}= 0\,,
	\label{eq:EOM}
\end{equation}
in which each term is actually a sum of the terms which have different
order of the factors, which is due to the symmetrized trace STr in
the Lagrangian, however such ``symmetrization'' is neglected in the
expression for brevity.
With our tachyon profiles the equation of motion becomes simple since
$T^\dagger T$ is proportional to the unit matrix and also $A_{\m\n}$
is diagonal.
Actually, eq.(\ref{eq:EOM}) with the tachyon profile (\ref{eq:TPd1})
becomes
\begin{equation}
 \left\{\frac{\hat\la u^2 f''V}{(1+\hat\la u^2f'^2)^2}
	- \frac{2fV'}{1+\hat\la u^2f'^2}
    -\frac{3\hat\la(f-ur f')V}{(1+\hat\la u^2f'^2)
	(r^2+\hat\la f^2)}\right\}\,\sqrt{-\det A}~M_{T}=0\,,
	\label{eq:EOM2}
\end{equation}
where
\begin{equation}
  \hat\la\equiv H_p^{\frac{1}{2}}\,\la\,,\quad
  M_{T}\equiv \mtx{cc}{
  \cos\t_1+i\sin\t_1\cos\t_2 & \sin\t_1\sin\t_2\,e^{i\t_3}\\
  -\sin\t_1\sin\t_2\,e^{-i\t_3} & \cos\t_1-i\sin\t_1\cos\t_2}\,,
\end{equation}
and
\begin{equation}
  -\det A =H_p^{-3}(1+\hat\la u^2f'^2)\,(r^2+\hat\la f^2)^3\,
	\sin^4\t_1\,\sin^2\t_2\,.
\end{equation}
Similarly, with the tachyon profiles, eqs.(\ref{eq:TPd1}),
(\ref{eq:TPd2}), (\ref{eq:TPd3}) and (\ref{eq:TPd4}), the equation of
motion (\ref{eq:EOM}) takes the same form as
\begin{equation}
 \left\{\frac{\hat\la u^2 f''V}{(1+\hat\la u^2f'^2)^2}
	- \frac{2fV'}{1+\hat\la u^2f'^2}
    -\frac{k\,\hat\la(f-ur f')V}{(1+\hat\la u^2f'^2)
	(r^2+\hat\la f^2)}\right\}\,\sqrt{-\det A}~M_{T}=0\,,
	\label{eq:EOM3}
\end{equation}
where
\begin{eqnarray}
  -\det A &=&H_p^{-3}\,(1+\hat\la u^2f'^2)\,
	(r^2+\hat\la f^2)^k\,
	\sin^{2[k-1]_+}\t_1\,\sin^{2[k-2]_+}\t_2\,,\\
  M_{T}&=&\frac{1}{f(ur)}~T\,,\\
  {}[n]_+ &\equiv& \left\{\begin{array}{cl} n & (n>0)\\
	0 & (n\le 0)\end{array}\right.\,,
\end{eqnarray}
and $k=3,2,1,0$ correspond to the profiles (\ref{eq:TPd1}),
(\ref{eq:TPd2}), (\ref{eq:TPd3}) and (\ref{eq:TPd4}), respectively.
In the $u\to\infty$ limit, the l.h.s.\ of eq.(\ref{eq:EOM3}) is
evaluated as
\begin{equation}
  \left\{ \frac{f''V}{u\hat\la^{1/2} {f'}^3}
	- \frac{2fV'}{u\hat\la^{1/2} f'}
    -\frac{k\hat\la^{1/2}(f-ur f')V}{u(r^2+\hat\la f^2)f'}
	\right\}\,(r^2+\hat\la f^2)^{k/2}
	\sin^{[k-1]_+}\t_1\,\sin^{[k-2]_+}\t_2\,M_T\,.
\end{equation}
This requires that $(Vf''f^k/{f'}^3)$ and $(V'f^{k+1}/f')$ should be
finite for any $r$ \cite{Sen:03},
\begin{equation}
  \left|\,\frac{f''(ur)f^k(ur)}{{f'}^3(ur)}~V(f^2(ur))\,\right|
	<+\infty\,,\quad
  \left|\,\frac{f^{k+1}(ur)}{f'(ur)}~V'(f^2(ur))\,\right|
	<+\infty\,,\label{eq:EOM12}
\end{equation}
and the third term in the braces implies
\begin{equation}
  \left|\,k\,\left(\frac{f(ur)}{f'(ur)} -ur\right)
	f^{k-2}(ur)\,V(f^2(ur))\,\right|
	<+\infty\,.\label{eq:EOM3rd}
\end{equation}
If the conditions (\ref{eq:EOM12}) and (\ref{eq:EOM3rd}) are
satisfied, then the tachyon profiles satisfy the equation of motion in
the $u\to\infty$ limit.
Let us examine these conditions further.
We can easily see that if we assume that $V$ damps faster than any
power of the arguments, eq.(\ref{eq:EOM12}) is satisfied.
This is also the case for $k\ge2$ in eq.(\ref{eq:EOM3rd}) and $k=0$
is, of course, trivial.
In the $k=1$ case, the above assumption for $V$ is adequate for
$r\ne0$, however, eq.(\ref{eq:EOM3rd}) is subtle in the $r\to0$
limit. Thus, we shall investigate the following condition in the
$r\to0$ limit,
\begin{equation}
  \left|\frac{f(ur)- ur f'(ur)}{f'(ur)f(ur)}\right| <+\infty\,.
	\quad(u\to\infty)\label{eq:cndf2}
\end{equation}
Due to the condition eq.(\ref{eq:cndf}), $f(x)$ can be
written (around $x=0$) by
\begin{equation}
  f(x) =\sum_{n=1}^K a_n\,x^n,\quad (a_1>0)
\end{equation}
where $K$ is some large number, or can be infinity.
Plugging into eq.(\ref{eq:cndf2}), we have
\begin{equation}
  \left|\frac{\sum\limits_{n=2}^K(1-n)a_n(ur)^n}{
	\sum\limits_{n=1}^{2K-1}
	(\sum\limits_{k=1}^n k\,a_k\,a_{n+1-k})
	\,(ur)^n}\right| <+\infty\,. \quad(u\to\infty)
\end{equation}
Hence we find that the condition (\ref{eq:EOM3rd}) is satisfied.

\section{Fluctuation around the vortex}\label{S:5}
In this section, we consider the small fluctuation around the tachyon
vortex representing the codimension 4 BPS D1-brane and give the
expected DBI action.
We set the gauge fields to be zero.
Following Ref.\cite{Sen:03}, we take the tachyon field with the
fluctuation fields $t^i(\xi)$ as
\begin{equation}
	T(x,\xi) = T(x^i-\la t^i(\xi))\,,
\end{equation}
where the coordinates along the D$p$-\aD{p} pairs are denoted by
$(\xi^\a,x^i)\hspace{1ex}(0\le\a\le (p-4),\ (p-3)\le i\le p)$, or
$\{\xi^\a\}$ are the
coordinates tangential to the vortex world-volume and $\{x^i\}$ are
perpendicular to them.
Also we include $(9-p)$ scalar fields which represent the fluctuation
along the transverse directions \footnote{We have neglected the
fluctuations of the relative motion of the branes.},
\begin{equation}
	\Phi^I(x,\xi) = y^I(\xi)\,,
\end{equation}
where $(p+1)\le I\le 9$. Then, from eqs.(\ref{eq:A}) and
(\ref{eq:Gmn}) we have
\begin{eqnarray}
  A_{ij} &=& H_p^{-\frac{1}{2}}\delta_{ij} + T_{ij}\,,\\
  A_{i\b} &=& -\la T_{ij}\,\p_\b t^j\,,\\
  A_{\a j} &=& -\la T_{ij}\,\p_\a t^i\,,\\
  A_{\a\b} &=& H_p^{-\frac{1}{2}}\eta_{\a\b}
	+\la^2H_p^{\frac{1}{2}} \p_\a y^I\p_\b y^I
	+\la^2 T_{ij}\,\p_\a t^i\,\p_\b t^j\,,
\end{eqnarray}
where
\begin{equation}
  T_{ij} = \frac{\la}{2}(\p_i T)^\dagger (\p_j T)
	+ \frac{\la}{2}(\p_j T)^\dagger (\p_i T)\,.
\end{equation}
To calculate the determinant, we define
\begin{eqnarray}
  &\hat{A}_{\a\n} = A_{\a\n} + \la A_{i\n} \p_\a t^i\,,&\quad
	\hat{A}_{i\n} = A_{i\n}\,,\\
  &\tilde{A}_{\m\b} = \hat{A}_{\m\b} +\la \hat{A}_{\m j}
	\p_\b t^j\,,&\quad \tilde{A}_{\m j} = \hat{A}_{\m j}\,,
\end{eqnarray}
which does not alter the determinant,
\begin{equation}
  \det{A} = \det{\hat{A}} = \det{\tilde{A}}\,.
\end{equation}
The components of $\tilde{A}$ are as follows,
\begin{eqnarray}
 \tilde{A}_{ij} &=& H_p^{-\frac{1}{2}}\,\delta_{ij} + T_{ij}\,,\\
 \tilde{A}_{i\b} &=& \la H_p^{-\frac{1}{2}}\,\p_\b t^i\,,\\
 \tilde{A}_{\a j} &=& \la H_p^{-\frac{1}{2}}\,\p_\a t^j\,,\\
 \tilde{A}_{\a\b} &=&  H_p^{-\frac{1}{2}}\,\left(\eta_{\a\b}
	+\la^2 \p_\a t^i\p_\b t^i
	+\la^2 H_p \p_\a y^I \p_\b y^I\right)\,.\label{eq:fl}
\end{eqnarray}
We shall define $\phi^I\equiv H_p^{\frac{1}{2}} y^I$. Then,
since we consider small fluctuations, we can regard $H_p$ as
a constant on the whole D-branes, eq.(\ref{eq:fl}) becomes
\begin{equation}
 \tilde{A}_{\a\b} \simeq  H_p^{-\frac{1}{2}}\,\left(\eta_{\a\b}
  +\la^2 \p_\a t^i \p_\b t^i + \la^2 \p_\a \phi^I \p_\b \phi^I \right)\,.
  \label{eq:fl2}
\end{equation}
Note that the fluctuations appear in eq.(\ref{eq:fl2}).
We may interpret both $\phi^I$ and $t^i$ as the transverse fields to
the brane.
Hence we extend easily the previous results and give the DBI action
including the fluctuations,
\begin{eqnarray}
 &&2\int\! d^6\xi\ \STr \left(e^{-\phi} V(|T|^2)\,
	\sqrt{-\det(P[G]_{\m\n}+2\pi\a'(\p_\m T)^\dagger(\p_\n T))}
	\right)\nn
 &&\to (2\pi\sqrt{\a'})^4 \left(2\int_0^\infty \!dy\ y^3\,V(y^2)\right)
    \int\!d^2\xi \sqrt{-\det\left(\eta_{\a\b}
	+\la^2 \p_\a \phi^I\,\p_\b \phi^I
	+\la^2\p_\a t^i\,\p_\b t^i\right)},\nn
\end{eqnarray}
where $0\le \a,\b \le 1,\ 2\le i \le 5$ and $6\le I \le 9$.
We can also add a gauge field, however, in that case the expression of
the tension will be changed \cite{Sen:03}.

\section{\boldmath BSFT in D$p$-brane background}\label{S:6}
In this section we shall analyze the system in BSFT.
Following Ref.\cite{BSFT:DDb}, we construct D$p'$-brane(s)
and a D$p'$-\aD{p'} system from two pairs
of D$p$-\aD{p} in the D$p$-brane background
($0<p-p'\leq4$).\footnote{Note that the two pairs of D$p$-\aD{p} as a
whole do not have RR-charges.}
The bulk action is given by
\begin{eqnarray}
 S_0&=&\frac{1}{4\pi\a'}\int_\Sigma d^2\sigma \sqrt{h}
    \left(\,h^{ab} \p_aX^\m \p_bX^\n
    +\a'\,\bar{\psi}^\m\rlap\slash\p\psi^\n\right)G_{\m\n}(X)\nn
  &&\qquad+\frac{1}{4\pi}\int_\Sigma d^2\sigma \sqrt{h} R\,\phi(X)
    +\frac{1}{2\pi}\int_{\p\Sigma}\!ds\,k_g\,\phi(X)\,,
\end{eqnarray}
where $G_{\m\n}$ and $e^{-\phi}$ are given in eq.(\ref{eq:Gmn}),
$h_{ab}$ is the metric of the worldsheet $\Sigma$ (disc), $R$ is
the worldsheet Ricci scalar and $k_g$ is the geodesic curvature.
Hereafter we ignore the transverse fluctuations, so that $G_{\m\n}$
and $\phi$ do not depend on the worldsheet coordinates and $\p_a
X^I=0$ for $I=p+1,\cdots,9$.
Since the Euler number $\chi$ of the disc is one and is given by
\begin{equation}
  \chi=\frac{1}{4\pi}\int_\Sigma d^2\sigma \sqrt{h}R
	+\frac{1}{2\pi}\int_{\p\Sigma} ds\,k_g=1\,,
\end{equation}
the bulk action in the conformal gauge is given by
\begin{equation}
 S_0=\frac{1}{4\pi}\int d^2z \left(\frac{2}{\a'}\,\p X^\mu\bar{\p}X^\n
    +\psi^\m \bar{\p} \psi^\n
    +\tilde{\psi}^\m \p \tilde{\psi}^\n \right)G_{\m\n} +\phi\,.
\end{equation}
We write the boundary values of $X^\m$ and $\psi^\m$ as
\begin{eqnarray}
 X^\mu(\tau)&=&X_0^\mu +\sqrt{\frac{\a'}{2}}\,
    \sum_{n=1}^\infty (X_n^\mu e^{in\tau}+X_{-n}^\mu e^{-in\tau})\,,\\
 \psi^\mu(\tau)&=&\sum_{r=1/2}^\infty
 (\psi_r^\mu e^{ir\tau} +\psi_{-r}^\mu e^{-ir\tau})\,,
\end{eqnarray}
where $\tau\ (0\le\tau <2\pi)$ parametrizes the boundary of the
worldsheet. To evaluate the bulk action with the boundary
values, we complexify $\tau$ by imposing the regularity as
$\im\tau\to\infty$\footnote{The worldsheet $\Sigma$ has
been mapped to the region of $0\leq \re z<2\pi,\,\im z\ge 0$.} and we
get
\begin{eqnarray}
 X^\mu(\tau,\bar{\tau})&=&X_0^\mu +\sqrt{\frac{\a'}{2}}\,
    \sum_{n=1}^\infty (X_n^\mu e^{in\tau}
	+X_{-n}^\mu e^{-in\bar{\tau}})\,,\\
 \psi^\mu(\tau,\bar{\tau})&=&\sum_{r=1/2}^\infty
 (\psi_r^\mu e^{ir\tau} +\psi_{-r}^\mu e^{-ir\bar{\tau}})\,.
\end{eqnarray}
Then the bulk action is calculated as
\begin{equation}
 S_0=\frac{1}{2}\sum_{n=1}^\infty n G_{\m\n}X_{-n}^\m X_n^\n
  +i\sum_{r=1/2}^\infty G_{\m\n}\psi_{-r}^\m \psi_r^\n +\phi\,.
\end{equation}

Let us consider the boundary interaction. The boundary term for the
tachyon is
\begin{equation}
 e^{-S_B}=\Tr \hat{P}\,e^{\int d\tau d\theta {\bf M}(\tau,\theta)}\,,
\end{equation}
where $\hat{P}$ stands for the path-ordered product in the superspace
$(\tau,\theta)$ and the $4\times4$ matrix ${\bf M}$ is given by
\begin{equation}
 {\bf M}=\mtx{cc}{0&T({\bf X})\\ T({\bf X})^\dagger &0}\,,
    \quad {\bf X}^\m=X^\m+i\sqrt{\a'}\theta\psi^\m.\label{eq:Min4}
\end{equation}
Here we consider only the tachyon and put the other fields to zero.
After some calculations, we can rewrite the boundary term
by using the ordinary path-ordering \cite{Mrcs}
\begin{equation}
 e^{-S_B}=\Tr P\exp\left[{\,\int\!d\tau (M_1-M_0^2)(\tau)}\right]\,,
\end{equation}
where ${\bf M}=M_0+\theta M_1$ and
\begin{equation}
  M_1-M_0^2=\mtx{cc}{-TT^\dagger & i\sqrt{\a'}\p_\m T\,\psi^\m \\
  i\sqrt{\a'}\p_\m T^\dagger\,\psi^\m & -T^\dagger T}\,.
\end{equation}

Now we consider the ABS construction \cite{ABS,W:98cd} of a BPS
D$(p-4)$-brane on the two pairs of D$p$-\aD{p} in the D$p$-brane
background.
The tachyon is given by
\begin{equation}
 \mtx{cc}{0&T\\T^\dagger&0}= u\sum_{k=p'+1}^p
	\Gamma_{k-p+4} X^k,\qquad (p-4\leq p' < p)\label{eq:ABS}
\end{equation}
where $\Gamma_i\ (i=1,2,3,4)$ are $4\times4$ matrices
\begin{equation}
  \Gamma_j = \mtx{cc}{0&\sigma^j\\\sigma^j&0}\,,\quad
  \Gamma_4 = \mtx{cc}{0&-iI\\ iI&0}\,,\quad(j=1,2,3)
\end{equation}
representing $SO(4)$ Clifford algebra,
\begin{equation}
 \{\Gamma_i,\Gamma_j\}=2\d_{ij}\,.
\end{equation}
In this case, the path-ordered trace is expressed by
using the boundary fermions,\footnote{Hereafter we omit the symbol of
summation over $k$.}
\begin{equation}
 e^{-S_B}=\int D\eta \exp\left[\int d\tau\left(
  \frac{1}{4}\dot{\eta}^k \eta^k
  -(uX^k)^2+i\sqrt{\a'}u\psi^k\eta^k\right)\right]\,.
\end{equation}
The equation of motion for $\eta$ gives
$\dot{\eta}^k=-i2\sqrt{\a'}u\psi^k$, and hence
\begin{equation}
 e^{-S_B}=\exp\left[\int d\tau\left(-u^2X^kX^k
	+\a'u^2\psi^k \p_\tau^{-1}\psi^k\right)\right]\,.
\end{equation}
Thus, the partition function is calculated as follows,
\begin{eqnarray}
 Z&=&\int DX D\psi\,e^{-S_0-S_B}\nn
  &=&\int\prod_{\m=0}^p \frac{dX_0^\m}{\sqrt{2\pi\a'}}
    \prod_{n=1}^\infty\frac{dX_{-n}^\m dX_n^\m}{4\pi}
    \prod_{r=1/2}^\infty[d\psi_{-r}^\m d\psi_r^\m]
    \,e^{-\phi}\,\exp\left(-2\pi u^2 X_0^k X_0^k \right)\nn
  && \times\exp\Biggl[-\sum_{n=1}^\infty \left(
    \frac{n}{2} G_{\m\n}X_{-n}^\m X_n^\n
    +2\pi\a' u^2 X_{-n}^k X_n^k \right)\nn
  &&\qquad -i\sum_{r=1/2}^\infty \left(G_{\m\n}\psi_{-r}^\m \psi_r^\n
	+\frac{4\pi\a' u^2}{r}\psi_{-r}^k \psi_r^k \right)\Biggr]\nn
 &=&(\sqrt{4\pi\a'}u)^{p'-p} H_p^{\frac{p-3}{4}}
  \left[\frac{\prod_{r=1/2}^\infty H_p^{-1/2}}{\prod_{n=1}^\infty
  nH_p^{-1/2}}\right]^{p'+1} \left[\frac{\prod_{r=1/2}^\infty
    \left( H_p^{-1/2}+\frac{4\pi\a' u^2}{r}\right)}
    {\prod_{n=1}^\infty (nH_p^{-1/2}+4\pi\a' u^2 )}\right]^{p-p'}
	\hspace{-2ex}\int\prod_{\mu=0}^{p'}
	\frac{dX_0^\m}{\sqrt{2\pi\a'}}\nn
 &=&(\sqrt{4\pi\a'}u)^{p'-p} H_p^{\frac{p-3}{4}}
  \left[\frac{1}{\sqrt{2\pi H_p^{1/2}}}\right]^{p'+1}
  \left[\frac{\cF (4\pi\a'u^2 H_p^{1/2})}{\sqrt{2\pi H_p^{1/2}}}\,
	\right]^{p-p'}
	\int\prod_{\mu=0}^{p'}\frac{dX_0^\m}{\sqrt{2\pi\a'}}\nn
 &=&\frac{H_p^{\frac{p-p'-4}{4}}}{(\sqrt{2\pi})^{p+1}}
    \left(\frac{\cF(4\pi\a' u^2H_p^{1/2})}{
	\sqrt{4\pi\a'u^2H_p^{1/2}}}\right)^{p-p'}
    \int\prod_{\mu=0}^{p'}\frac{dX_0^\mu}{\sqrt{2\pi\a'}}\,,
\end{eqnarray}
where we have used the $\zeta$-function regularization,
\begin{equation}
 \frac{\DSP\prod_{r=1/2}^\infty \left( a+\frac{x}{r} \right)}
      {\DSP\prod_{n=1}^\infty (na+x)}
  = \sqrt{\frac{a}{2}}\,
    \frac{\Gamma(\frac{x}{a}+1)}{\Gamma(\frac{x}{a}+\frac{1}{2})}
  = \frac{x\,4^{x/a}\,\Gamma(x/a)^2}{2\sqrt{2\pi a}\,\Gamma(2x/a)}\,,
\end{equation} and $\cF$ is defined by
\begin{equation}
 \cF(x)=\sqrt{2\pi}~\frac{\prod_{r=1/2}^\infty%
	 \left(1+\frac{x}{r}\right)}{\prod_{n=1}^\infty (n+x)}
  =\frac{x\,4^x\, \Gamma(x)^2}{2\Gamma(2x)}\,.
\end{equation}
Therefore we obtain the spacetime string field action,
\begin{equation}
 S=Z= 4\,T_p\, (\sqrt{2\pi\a'})^{p-p'}\,H_p^{\frac{p-p'-4}{4}}
    \left(\frac{\cF(4\pi\a' u^2H_p^{1/2})}{
	\sqrt{4\pi\a'u^2H_p^{1/2}}}\right)^{p-p'}
    \int\prod_{\mu=0}^{p'}dX_0^\mu\,,\label{eq:4dact}
\end{equation}
where we have fixed the overall constant by hand \cite{BSFT:DDb},
indicating that the original system is the two pairs of
D$p$-\aD{p} in the D$p$-brane background.

By the tachyon condensation of taking the $u\to\infty$ limit,
the lower-dimensional brane will be produced.
In the asymptotic region, $\cF(x)$ behaves as
\begin{equation}
 \cF(x)\sim \sqrt{\pi x}\,.\quad(x\to\infty)
\end{equation}
Thus in the $u\to\infty$ limit, we obtain
\begin{eqnarray}
 S&\to& (\sqrt{2})^{4-p+p'} (2\pi\sqrt{\a'})^{p-p'} T_p
	H_p^{\frac{p-p'-4}{4}}\int\prod_{\mu=0}^{p'}dX_0^\mu\nn
 &=&(\sqrt{2})^{4-p+p'} T_{p'} H_p^{\frac{p-p'-4}{4}}
	\int\prod_{\mu=0}^{p'}dX_0^\mu\,,
\end{eqnarray}
where
\begin{equation}
    T_{p'} = (2\pi\sqrt{\a'})^{p-p'} T_p\,.
\end{equation}
This indicates one BPS D$(p-4)$ when $p'=p-4$, a non-BPS D$(p-3)$ when
$p'=p-3$, a pair of BPS D($p-2$)-\aD{(p-2)} when $p'=p-2$ and two
non-BPS D$(p-1)$-branes when $p'=p-1$, respectively. And only the
BPS D$(p-4)$-brane is stable in the D$p$-brane background because
$H_p$ disappears (cf.\ Ref.\cite{KLP}).
Actually, these non-BPS D-branes and a pair of D\aD{} branes are
understood once we consider the Chern-Simons couplings between
the branes and the RR background. The relevant partition function is
given by  \cite{BSFT:DDb}
\begin{equation}
   Z_{RR}\propto \int C\wedge \Str e^{2\pi i{{\cF}_T}}\,,
\end{equation}
where $C$ is the RR potential, $\cF_T$ is the curvature of the
superconnection \cite{Q,BGV},
\begin{equation}
  i{\cF}_T=\mtx{cc}{-TT^\dagger& i\sqrt{\a'}\,dT\\
	i\sqrt{\a'}\,dT^\dagger&-T^\dagger T}\,,
\end{equation}
and the supertrace is defined by
\begin{equation}
  \Str M = \Tr (-)^F\,M = \Tr \left[\mtx{cc}{1&0\\0&-1}\,M\right].
\end{equation}
Thus, by using the tachyon in eq.(\ref{eq:ABS}) with $X^k$ being only
zero modes, we have
\begin{equation}
  Z_{RR}\propto \int C\wedge \Str\Biggl[
	\prod_{k=p'+1}^p e^{-2\pi u^2 (x^k)^2}(2\pi\sqrt{\a'}\,
	\Gamma_{k-p+4}\,dx^k)\Biggr]\,.
\end{equation}
Then, $Z_{RR}$ vanishes when $(p-3)\leq p'\leq(p-1)$, which means that the
resultant brane(s) has no RR-charge. On the other hand, $Z_{RR}$
is non-vanishing when $p'=p-4$, indicating a BPS D-brane.

So far we have investigated two pairs of D$p$-\aD{p} in the D$p$-brane
background. We can easily see that one pair of D$p$-\aD{p} in
the D$p$-brane background leads to a similar equation
\begin{equation}
 Z_{D\bar{D}}= 2\,T_p\, (\sqrt{2\pi\a'})^{p-p'}\,H_p^{\frac{p-p'-4}{4}}
    \left(\frac{\cF(4\pi\a' u^2H_p^{1/2})}{
	\sqrt{4\pi\a'u^2H_p^{1/2}}}\right)^{p-p'}\hspace{-2ex}
    \int\prod_{\mu=0}^{p'}dX_0^\mu\,,\quad(p-2\leq p'<p)
\end{equation}
once we take ${\bf M}$ in the boundary interaction as a $2\times 2$
matrix similar to eq.(\ref{eq:Min4}) and the tachyon as
\begin{equation}
 \mtx{cc}{0&T\\T^\dagger&0}=
	u\sum_{k=p'+1}^p \sigma^{k-p+2}X^k\,.\qquad (p-2\leq p' < p)
\end{equation}
When $p'=p-2$, one BPS D$(p-2)$-brane can be produced in the
$u\to\infty$ limit
\cite{BSFT:DDb} and in that case $Z_{D\bar{D}}$ becomes
\begin{equation}
 Z_{D\bar{D}}\to T_{p-2} H_p^{-\frac{1}{2}}
	\int\prod_{\mu=0}^{p-2}dX_0^\mu\,,\label{eq:1DDb}
\end{equation}
where $H_p$ does not disappear. This shows that the resultant
D$(p-2)$-brane feel attractive force from the background of $N$
D$p$-branes, which is consistent with the well-known fact that
the D$p$-D$(p-2)$ system is unstable.

\section{Conclusion and discussion}
In this paper we have considered the tachyon condensation in the
unstable D-brane systems.
For the system of two pairs of D5-\aD{5} in the background of large
$N$ D5-branes (see Figure \ref{F:2}), we considered the
DBI effective action and gave the explicit forms of the tachyon
profile and we have shown that one of the profiles does lead to one
BPS D1-brane in the D5-brane background, which is expected to be a
stable endpoint.

We have also considered the same system in the BSFT framework.
To lift the DBI effective action in the D$p$-brane
background to BSFT,
we studied the worldsheet sigma-model action in this background
and obtained the reasonable $H_p$ dependence in the actions of the
lower-dimensional D-brane(s) from the ABS construction.
Furthermore, we have shown that one BPS D3-brane from a pair of
D5-\aD5 gets force from $N$ D5-branes, which is different from the
D1-brane (codimension 4) case.

Some comments are in order.
In our starting configuration, two pairs of D5-\aD{5} will get
forces equally from all the $N$ D5-branes. This implies that there
also exists attractive force between the pairs of the D5-\aD{5} and
$N$ D5-branes whose distance is of $O(\sqrt{\a'})$.
This would be consistent with the fact that there are tachyonic modes
in the strings between the two \aD5-branes and the $N$ D5-branes if
all the D-branes coincide or close within the distance of
$\sqrt{\a'}$.
On the other hand, our result of one BPS D1-brane in the D5-brane
background, which is stable, implies that we could follow the
principle of superposition, that is, all the D5-branes except two seem
to be spectators.
This suggests that if we consider the coincident $(N+2)$ D5 and two
\aD5-branes, the tachyons coming from the strings between the $N$ D5
and the two \aD5-branes are no longer tachyonic after the
condensation.
On the other hand, the action of the resultant BPS D3-brane in
eq.(\ref{eq:1DDb}) is proportional to the negative power of $H_5$ and
it suggests that the tachyons from $N$ D5-branes and a \aD5-brane
remain tachyonic after the condensation. These correspond to the fact
that the NS ground state of the string between a D$p$-brane and
D$p'$-brane is massless when $p-p'=4$ and tachyonic when $p-p'=2$.

We considered the static solutions in this paper.
It is, of course, important to analyze a time dependent solution
in our unstable system.
If two pairs of D5-\aD5 are located far from the stack of $N$
D5-branes, the forces from $N$ D5-branes are small and hence
it may be possible to construct the solution which represents
the decay into a BPS D1-brane.
Since the total energy should be conserved in the decay process,
it is plausible to expect that the resultant system is a lower
dimensional D-brane surrounded by the tachyon matter as in
Ref.\cite{IU}.
However, the stack of $N$ D5-branes would experience attractive force
from the tachyon matter and the BPS D1-brane would also feel
attractive force from the tachyon matter. Therefore the final state
might be a gravitationally bounded state of D1-brane, D5-branes and
the tachyon matter.

\vspace{10pt}
\noindent\textbf{Note added:}
While we were completing the manuscript, we became aware of
Ref.\cite{HT}, in which it is pointed out that the tachyon
condensation on $(N+2k)$D4-$2k$\aD4 to $N$D4-$k$D0 explains the ADHM
construction of instantons.

\vspace{10pt}
\noindent
\textbf{Acknowledgments:}
This work is supported in part by
the Postdoctoral Research Program of Sungkyunkwan University (2005),
and is the result of research activities (Astrophysical
Research Center for the Structure and Evolution of the Cosmos
(ARCSEC)) supported by Korea Science \& Engineering Foundation
(A.I.).
It is also supported in part by MEXT Grant-in-Aid for
the Scientific Research \#13135212 (S.U.).


\end{document}